# Randomness: what is it and why does it matter?

Mario Stipčević[1,*]

[1] *Photonics and Quantum Optics Research Unit, Center of Excellence for Advanced Materials and Sensing Devices, Ruđer Bošković Institute, Bijenička cesta 54, 10000 Zagreb, Croatia*
*Corresponding author: mario.stipcevic@irb.hr*

**Abstract** – Randomness is a crucial resource for a broad range of important applications, such as Monte Carlo simulation and computation, generative artificial intelligence and cryptography. But what is randomness? A widely accepted definition has eluded researchers thus far, yet without one any work that relies on notion of randomness lacks scientific rigor and its results are questionable. Here, I propose an information-theory-based definition of randomness which, unlike previous attempts, does not list desired properties of the generated number sequence, but rather focuses on the physical process of random number generation itself. This approach results in a definition which complies with our intuitive perception of randomness. It is demonstrated to be non-empty and verifiable. Moreover, a new quantity named "randomness deviation" allows for a practical measure of quality of a random number generating process or a device. An expression for it is derived for a Markovian process, which is frequently encountered in practice. Finally, a process-oriented definition of a random number sequence completes the toolbox needed for understanding, proving, and practical use of randomness.

**Subject** - Interdisciplinary Physics

## 1. Introduction

Randomness is essential in many areas of research and application, some of which require provability, such as: cryptography, stochastic computing, randomized algorithms, Monte Carlo simulations and generative AI. Notably, cryptographic protocols require the generation and use of secret random numbers that must be unknown to attackers even in principle which necessitates proof of their randomness. However, so far, there is no widely accepted definition of a random number sequence, or randomness in general, what would enable such a proof.

### 1.1. Early definitions of randomness

One of the first attempts to define a random number sequence in modern mathematical formalism was that of von Mises developed from 1919-1933 [1] where the definition is based on two axioms: the first stating equality of probabilities of ones and zeros and the second requiring "irregularity" or "unpredictability". For example a sequence of all zeros is not random because it lacks ones, while two alternating sequences 01010101... and 001100110011... are not random by being easily predictable even though they have an equal probability of zeros and ones. However, his attempts to build an axiomatic mathematical theory of randomness led to inconsistencies which von Mises was able to evade for decades by adjusting the theory, however in 1939 Ville has shown that it is possible to construct biased sequences that comply with von Misses' axioms [1], which ended that program. The legacy of von Mises is the idea that if a sequence is to be called random, out of all possible binary combinations of the same length some must be omitted due to some criterion.

A more successful definition of randomness was presented in the paper titled "The definition of randomness" by Martin-Löf in 1966 [2]. He was a PhD student of famous mathematician Andrei Kolmogorov who happened to be working, at the time, on the question of "What is the simplest algorithm that would reproduce a given finite sequence of symbols?", or in the modern jargon, on the lossless compression. Kolmogorov defined a measure known today as "Kolmogorov complexity". The idea lingering in the air was to define a random sequence as the one whose complexity, measured in bits, is roughly equal to its length. Inspired by the work of his mentor, as well as accepting the aforementioned legacy of von Misses which he studied in detail [1], Martin-Löf dreamt of a definition of a random number sequence as an entity that is incompressible. The idea behind that was that if a block of data can be losslessly compressed to a sequence of a lesser length, then the information contained in the extra length is calculable form the smaller lot (by means of decompression) and therefore the original sequence is not entirely random. In another words, a sequence (vector) of bits $x = (x_0, x_1, x_2, ...)$ is considered random if there exists *no* algorithm that outputs $x$ and whose length is shorter than the length of $x$. Getting rid of logical inconsistencies of the definition came at a price: it only works for infinite sequences. Over time, the identical definition has been reached from three different requirements: incompressibility, patternlessness and unpredictability [3] and the unique name "1-randomness" has been coined for it. However, checking the compliance of a sequence of numbers to the definition implies applying all possible compression algorithms whose description requires lesser or equal number of symbols than the description of the sequence being evaluated, and checking whether any of them succeeds in compressing the sequence. Neither Martin-Löf, nor anyone else up to date, has provided a workable algorithm for checking out compliance of a sequence to the definition, nor a measure for quantifying randomness. On the contrary, Wolfram finds out that "it turns out to be *undecidable* in general whether any particular sequence is random" because at least some of algorithms will not halt [4]. Therefore in spite of being conceptually intriguing, this definition has only a theoretical value.

## 1.2. Definition of finite random number sequences

Problem with the definition of a finite random sequence is that all possible sequences of the same length should be equally probable and therefore equally random, which basically means that randomness has no meaning. On the other hand, under the provision that only finite random sequences are of merit for any practical use, finding an appropriate definition is of vital importance.

In one of the most comprehensive computer engineering surveys on randomness of its time, first published in 1969 [5], Donald Knuth quests for a definition of an infinite random number sequence by performing a meticulous deduction of six increasingly improved candidate definitions. The last, R6, essentially concurs with 1-randomness.

But, Knuth goes further, into uncharted territory, by constructing two candidates for the definition of finite random sequences, which he names Q1 and Q2. While Q1 depends on the notion of k-randomness - the closest cousin of 1-randomness that could be applied to finite sequences, Q2 only depends on the balance of the frequencies of 0s and 1s being equal within certain arbitrary level $\epsilon$. The definition Q1 applied to 11-bit long sequence, it declares that non-random are 178 out of 2048 possible sequences, namely those that contain 7 or more consecutive 0s or 1s, the sequence 01010101010 and its binary complement. But that leads to a problem. Namely, an obvious intuitive requirements on random sequences is that concatenation of any number of random sequences should result in a random sequence. But the "expelled" sequences will be missing no matter how long the concatenated sequence is! At this point, one might be tempted to accept the desperate proposal of Church to abandon the idea of defining randomness altogether [6]!

## 1.3. Pseudorandom sequences

Pseudorandomess is a well-known principle of generating deterministic sequence of numbers by a suitable mathematical function such that they possess certain statistical properties desirable of true random sequences. For this to work, one needs to supply an initial vector for the argument, so called *seed*. Pseudorandom sequences are not random, if anything, because they are highly compressible by design. In Knuth's words: "Pseudorandom generally means not random, but it *appears* random" [5]. A glance into the well-developed theory of pseudorandom number generators can be found in Refs. [7] and [8].

Stephen Wolfram generalized pseudorandomness by substituting the generating functions with *elementary cellular automata* algorithms, also known as "Games of life". In his seminal work "New kind of science" published in 2002 [4] (last edition in 2019) Wolfram states that cellular automata govern physical laws, but are seeded by a small amount of true randomness from the Nature which ensures they never repeat exactly. While the physical theory has been widely criticized for its arbitrariness, Wolfram correctly deduces that no randomness can be created or amplified by deterministic algorithms, a point that we will return to later. Curiously, in his book the word "randomness" is mentioned over 550 times, yet it is never defined.

## 1.4. Physical randomness

Most of the recent research studies on randomness agree that a sequence of random numbers should be generated by a physical process which is by itself considered random [9, 10]. Such a process may rely on electronics noise [11], chaos in lasers [12] etc, but preferably on the fundamental quantum randomness, such as atomic or nuclear decay, photoelectric effect [13], random phase fluctuations in lasers etc. State of the art is to generate a bunch of imperfect random numbers, estimate how much random bits it is worth, and then use techniques of randomness extraction [14] and (optionally) randomness certification [15] to arrive to a smaller set of almost perfect random numbers. The final step consists in verifying that a random number sequence passes comprehensive statistical tests, such as: NIST Statistical Test Suite (STS) [16], TestU01 suite [17], DIEHARDER suite based on the work of G. Marsaglia [18] and a set of efficient tests intended for on-the-fly monitoring of sanity of a physical random number generator (RNG) such as NIST's SP800-90B [19].

However, proof of randomness through testing it is not all that simple because there are arguably infinitely many randomness tests and, just as in the Martin-Löf case, some of them may be undecidable, or simply unknown. Consequently, there is a wide consensus that passing randomness test(s) does not prove randomness of the tested number sequence: it merely means that it is good enough for applications for which the tested statistical properties are relevant and sufficient. Notably, some pseudorandom sequences may pass statistical tests quite well, even though they are clearly not random.

The discussion thus far has shown insurmountable problems with defining a *random number sequence*. Therefore, in this work, I define a *physical random number generator* capable of generating random numbers. This paradigm shift will allow us to arrive to an intuitive and applicable definition of randomness that will remove arbitrariness in understanding and operational use of that important concept.

## 2. Physical definition of randomness

Let us consider the simplest atom of all, the hydrogen atom, prepared in the 3p state. The state is unstable and will transition either to 1s state with probability of ≈ 0.875 emitting a photon of wavelength 102.6 nm (Lyman series), or to 2s state with probability of ≈ 0.125 emitting a photon of 656.3 nm (Balmer series). Quantum mechanics can accurately calculate the decay probabilities, but it *cannot* tell us which of the two decays will be observed in an actual measurement: not because it is incomplete or inaccurate, but because it is fundamentally impossible to predict outcome of that process. In more technical terms, according to the quantum mechanics, there is no placeholder for information in an atom that could allow one to either set or predict which decay path will take place. It is this fundamental *absence of information* on the next decay and at the same time the *necessity of the atom to decay*, which combined cause the randomness in this process. This random doing of Nature we cannot predict, influence nor deny its existence. Randomness simply *is* out there, in the Nature itself, and that is where we are going to search for its definition.

Instead of an atom, which has fixed transition probabilities, in the rest of the discussion we will consider a beamsplitter based quantum random number generator (QRNG), shown in Fig. 1, which has an advantage that its probabilities of generating bit values of 0 or 1 can be, at least in theory, tailored such as to be equal. I want to stress that the QRNG itself *not* the subject of this study: this particular QRNG setup is well-known in the literature [13] and is considered here in an idealized form with the sole purpose of arriving to the definition of randomness.

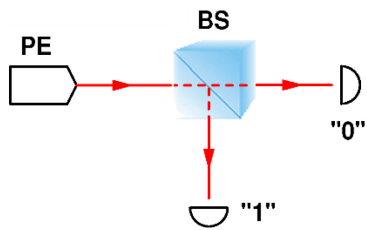

**Figure 1**. An idealized, beamsplitter-based quantum random number generator: the photon emitter PE sources a photon. The photon enters the perfectly balanced beamsplitter BS which either transmits or reflects the photon with an equal probability. The photon ends up being absorbed and detected by a detector in its path, thus generating a binary number of value 0 or 1.

Let us assume a single photon is entering the non-polarizing beamsplitter BS in Fig. 1. The state of the photon can be seen as a linear superposition of two orthogonal pure quantum states denoted $|0\rangle$ and $|1\rangle$:

$$|\Psi\rangle = \alpha|0\rangle + \beta|1\rangle \quad (1)$$

where $\alpha^2$ and $\beta^2$, satisfying relation $\alpha^2 + \beta^2 = 1$, are the probabilities of the photon hitting the two detectors denoted as "0" and "1", interpreted as generation of a binary value 0 or 1 respectively. Analogously to the hydrogen atom example discussed above, quantum mechanics, in fact the Nature, allows us only to predict outcome probabilities $p(0) = \alpha^2$ of the bit value 0 and $p(1) = \beta^2$ of the bit value 1, but not the outcome of any particular measurement. This fundamental impossibility of prediction is crucial for understanding the concept randomness: there is no placeholder for information in this whole physical system that could determine, or even hint, the outcome of the future measurement and *that* is what makes the outcome random.

The photon emitter PE in Fig. 1 could, for example, be an atom which was previously prepared in an appropriate unstable excited state and which has only one possible decay route to the ground state. After emission of a photon, the atom is exhausted and there are no more emissions. But, if we "reset" it to the same excited state it will "fire" again. Since after the "reset" the atom is, according to the laws of quantum mechanics, in exactly the same excited state as it was before, the decay photon will again choose one of the two spatial paths in a way that we perceive as "random", because it has no reference to anything that happened before. More precisely, "random" here means that the previous experiment(s) left no trace in the setup that could affect future measurements *and* that the *bias*, defined as:

$$b = p(1) - p(0) \quad (3)$$

is equal to zero. This perception of randomness is strictly formulated by the following definition.

**Definition D1**. *Random bit generator is a physical device that produces a sequence of classical bits $x_1$, $x_2$, $x_3$, ... in such a way that a bit value being generated does not contain any information on any subset of bits previously generated by the device and that the a priori expected value of a bit is equal to ½, that is:*

$$I(x_{i+1}; \{x_1, \ldots, x_i\}) = 0; \quad (4)$$

$$E(x_{i+1}) = \frac{1}{2}. \quad (5)$$

For shortness, in the rest of the paper we will denote the set of all previously generated bits as $X_p = \{x_1, \ldots, x_i\}$, thus the first condition may be concisely written as $I(x; X_p) = 0$, and second as $E(x) = 1/2$, where $x$ means "any generated bit". Even though it might seem that the definition requires knowledge of the past of the generator, it does not. It could be interpreted such that the process of generating a number makes no influence on the number(s) that will be generated in the future, with respect to the two requirements.

There are two novelties in this definition with regard to the state-of-the-art: 1) *what* is defined: instead of defining a mathematical abstraction (a random sequence of bits) defined is a physical device (a random number generator); and 2) *how* it is defined: instead of defining what randomness *is* (e.g. by listing a set of tests numbers must pass) we define what randomness *isn't* (any mutual information among bits and any bias).

For a definition to be valid, it must define a non-empty set and must be verifiable. It is fascinating that quantum mechanics permits existence of such a device, and we have just discussed one: the beamsplitter QRNG, shown in Fig. 1! Therefore the definition is not empty. It will be demonstrated that it is verifiable too.

Definition **D1** basically divides any possible *deviations* from the perfect randomness in two classes: correlations (described by Eq. (4)) and bias (described by Eq. (5)). Note that in order to generate correlated bits, a RNG must have some sort of memory so to allow for bit values generated at different instances (times) to influence each other. If a generator does not allow bit values to influence each other (which is true in the above idealized beamsplitter) then the bits produced by the generator are statistically independent. This however still does not guarantee the maximal randomness because bit values can be biased. Bias does not require interaction among bits: it can be imprinted into the generator itself (for example using a beamsplitter with the uneven splitting probabilities would cause un-even probabilities of 0's and 1's). In either of the two limiting cases of bias, namely $p(0) \to 0$ or $p(1) \to 0$, the random number generating process becomes deterministic. Let us illustrate the use of this definition on two elucidating examples.

**2.1. Imperfect RNG example 1: beamsplitter-based QRNG with detectors that exhibit dead time**

Definition **D1** is powerful because it allows one to actually check whether a generator satisfies it and even to determine a numerical measure of deviation from randomness. As an illustrative example of that, let us consider a beamsplitter QRNG (shown in Fig. 1) in which two identical photon detectors ("0", "1") are imperfect in the sense that they each feature the dead time $\tau_d > 0$. Let us further suppose that PE is a light emitting diode (LED). Photons from such a source are detected as random Poisson events, with inter-arrival (waiting) times that follow the exponential probability density function [13]:

$$\frac{1}{\tau} e^{-t/\tau} \tag{6}$$

where $\tau$ is the average photon inter-arrival time. When a photon is detected by a detector, the detector will become insensitive for next $\tau_d$ and if the next photon arrives during that time it can only be detected by the other detector. This effect will cause an elevated probability of generation of 1 immediately after 0 and 0 immediately after 1 thus causing a negative serial autocorrelation, even though it does not change bias. The serial autocorrelation coefficient of lag $k$ is generally defined as [5]:

$$a_k = \frac{\sum_{i=1}^{N-k}(x_i - \bar{x})(x_{i+k} - \bar{x})}{\sum_{i=1}^{N-k}(x_i - \bar{x})^2} \tag{7}$$

where $\{x_1 \ldots x_N\}$ in as $N$ bits long random sequence. When $a_k$ is estimated using Eq. (7) *a posteriori*, from a set of generated bits, its statistical uncertainty is $\sigma_a = 1/\sqrt{N}$. In our discussion, however, $N$ is irrelevant (we can imagine that $N \to \infty$) because we do not regard $a_k$ as a result of an *a posteriori* measurement, but as a characterizing parameter of the random number generator.

We assume that BS is balanced and that both detectors have identical dead time, therefore, because of symmetry, there is no bias i.e. $\bar{x} = 1/2$. By considering the four possible combinations for pair $(x_i, x_{i+1})$ in Eq. (7) it can be shown that $a_1$ is equal to:

$$a_1 = 2\,\mathbb{P}(x_{i+1} = x_i) - 1 \tag{8}$$

where $\mathbb{P}(x_{i+1} = x_i)$ is a probability that bit $x_{i+1}$ is equal to the previous bit $x_i$. First, there is a probability of 1/2 that the photon will chose the path towards the same detector as the previous photon did. This particular detector, because it was hit previously, is dead for $\tau_d$ so the overall detection probability is:

$$\mathbb{P}(x_{i+1} = x_i) = \frac{1}{2}\left(\frac{1}{\tau}\int_{\tau_d}^{\infty} e^{-t/\tau} dt\right) = \frac{1}{2} e^{-\tau_d/\tau} \approx \frac{1}{2}\left(1 - \frac{\tau_d}{\tau}\right). \tag{9}$$

where the numerical approximation in the last step, as well as neglecting a small probability that both detectors might have been left dead from even earlier detections, are both valid in the limit $\tau_d \ll \tau$. In a realistic case, the mean photon emission rate per detector could be 1 MHz ($\tau = 1000$ ns) and dead time could be $\tau_d = 40$ ns, leading to $a_1 \approx -0.04$: a deviation from randomness large enough to be statistically significant for a sequence of only a few hundred bits, even though *a priori* probabilities of generating zeros and ones are the same, that is $b = 0$. If $a_1$ is small enough then correlation between non immediate neighboring bits is negligible and the first term in **D1** can be approximated by [20]:

$$I(x_{i+1}; x_i) = \sum_{x_{j+1} \in \{0,1\}} \sum_{x_j \in \{0,1\}} p(x_{j+1}, x_j) \log_2 \frac{p(x_{j+1}, x_j)}{p(x_{j+1})p(x_j)} \tag{10}$$

where $p(x_{i+1}, x_i)$ and $p(x_i)$ are the *joint* and the *marginal* distributions respectively, and the summation goes over all four combinations of bits: $(x_{j+1}, x_j) \in \{(0,0)(1,1)(0,1)(1,0)\}$. The equation describes the mutual information exactly if bit generation is a Markov process because then, by definition, $I(x_{i+1}; X_p) = I(x_{i+1}; x_i)$, which is characterized by the exponential fall-off of autocorrelation coefficient with the lag: $a_k = a_1^k$ [21]. Markov process is a good approximation for physical generators in which an intentional lack of memory ensures that only adjacent bits may be non-negligibly correlated, usually due to imperfection(s) in the actual realization of a theoretically perfect physical RNG. Taking into account Eq. (8), joint and marginal distributions, for our exemplary imperfect QRNG, are summarized in Table 1.

| $x_j$ \ $x_{j+1}$ | 0 | 1 | $p(x_j)$ |
|---|---|---|---|
| 0 | $(1+a_1)/4$ | $(1-a_1)/4$ | 1/2 |
| 1 | $(1-a_1)/4$ | $(1+a_1)/4$ | 1/2 |
| $p(x_{j+1})$ | 1/2 | 1/2 | |

**Table 1**. Joint distribution (middle of the table) and marginal distributions (rightmost column and lowest row) distribution for the QRNG with detectors that exhibit a dead time.

Inserting values in Table 1 into Eq. (10) yields:

$$I(x_{i+1}; x_i) = \frac{1}{2}(1+a_1)\log_2(1+a_1) + \frac{1}{2}(1-a_1)\log_2(1-a_1). \tag{11}$$

Assuming that any correlations among non-neighboring bits are negligible, that is $I(x_{i+1}; X_p) \approx I(x_{i+1}; x_i)$, and expanding the right side of Eq. (11) into Taylor series, one obtains:

$$I(x; X_p) \approx \frac{a_1^2}{2\ln(2)} + O(a_1^4) \tag{12}$$

where $x$ denotes any generated bit and $X_p$ denotes all bits generated before it. A comparison of this parabolic approximation with the exact function of Eq. (11) is shown in Fig 2.

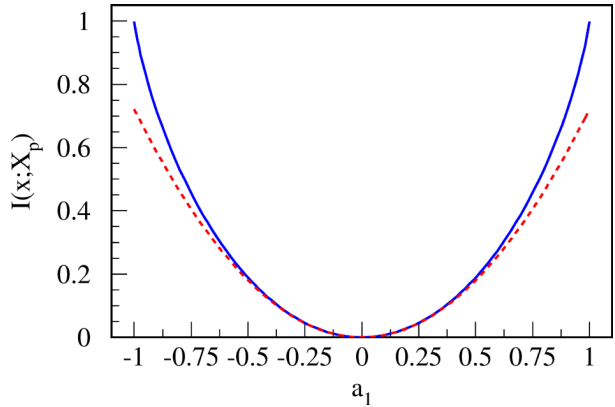

**Figure 2**. Common information as a function of serial autocorrelation coefficient with lag 1 for Markov bit generating process: exact (full blue line) and the parabolic approximation of Eq. (12) (dashed red line).

By establishing the non-vanishing autocorrelation of this RNG, we have proven that, according to **D1**, it is *not* random and moreover we are able to establish the nature of non-randomness and estimate its magnitude.

### 2.2. Imperfect RNG example 2: a QRNG with an unbalanced beamsplitter and perfect detectors

Now let us consider yet another example of a beamsplitter QRNG, in which detectors exhibit no dead time, but the beamsplitter divides beam power unevenly such that detectors "1" and "0" detect with bias $b$. Taking into account that, from definition in Eq. (3) it follows that $p(0) = (1-b)/2$ and $p(1) = (1+b)/2$, as well as that bits are not correlated, joint and marginal distributions are as shown in Table 2.

| $x_j$ \ $x_{j+1}$ | 0 | 1 | $p(x_j)$ |
|---|---|---|---|
| 0 | $(1-b)^2/4$ | $(1-b^2)/4$ | $(1-b)/2$ |
| 1 | $(1-b^2)/4$ | $(1+b)^2/4$ | $(1+b)/2$ |
| $p(x_{j+1})$ | $(1-b)/2$ | $(1+b)/2$ | |

**Table 2**. Joint distribution (middle of the table) and marginal distributions (rightmost column and lowest row) distribution for the QRNG with un-balanced beamsplitter.

Substitution of values in Table 2 into Eq. (10) yields:

$$I(x_{i+1}; x_i) = (1-b)^2 \log_2 \frac{(1-b)^2}{(1-b)^2} + 2\log_2 \frac{1-b^2}{1-b^2} + (1+b)^2 \log_2 \frac{(1+b)^2}{(1+b)^2} \equiv 0. \quad (13)$$

Note that this relation holds generally: bias does not contribute to mutual information and therefore, conversely, common information does not contribute to bias. By induction, we conclude $I(x; X_p) = 0$ for uncorrelated generation of bits. Having in mind the following identity for Shannon's entropy [20]:

$$H(x) = H(x|X_p) + I(x; X_p) \quad (14)$$

and the well-known logarithmic expression for Shannon's entropy in the case of uncorrelated bits, we arrive to:

$$H(x) = H(x|X_p) = -[p_0 \log_2(p_0) + p_1 \log_2(p_1)]$$
$$= -\frac{1+b}{2}\log_2\frac{1+b}{2} - \frac{1-b}{2}\log_2\frac{1-b}{2}. \quad (15)$$

Expansion into Taylor series for $|b| \to 0$ yields:

$$H(x|X_p) = 1 - \frac{b^2}{2\ln(2)} + O(b^4). \quad (16)$$

Now, inserting Eqs. (12) and (16) into Eq. (14) we get an expansion of the overall entropy for a RNG operating as a Markov process:

$$H(x) \approx 1 - \frac{a_1^2 + b^2}{2\ln(2)}. \quad (17)$$

## 3. Deviation from randomness

The special-case result in Eq. (17) for the Markov process, motivates to define a numerical measure for *deviation from randomness* for a general random number generating process, as follows.

**Definition D2**. *Deviation from randomness, D, of a random bit generator is defined as:*

$$\mathcal{D} = 1 - H(x|X_p) - I(x; X_p) \quad (18)$$

*where $x$ is a bit produced by the generator and $X_p$ is a set of all bits generated previously by the generator.*

Deviation $\mathcal{D}$ spans from 0 (for a perfect random generating process) to 1 (for a deterministic generating process). It can be used as a unique criterion for the quality of randomness of a physical RNG. For the Markov process, we readily obtain an approximate expression form Eq. (17):

$$\mathcal{D} \approx \frac{a_1^2 + b^2}{2\ln(2)}. \quad (19)$$

Using a computer program for simulation of the binomial random bit generation process with an arbitrary bias and arbitrary serial autocorrelation [22], and evaluating Shannon entropy with the help of the public program ENT [23], it has been found that the error of Eq. (19) is less than 0.25% if $|a_1| \leq 0.1$ and $|b| \leq 0.1$. Because such a large amounts of bias and autocorrelation would render the generator completely useless anyway, it could be safely concluded that the expression given in Eq. (19) is adequate for virtually all practical purposes.

A practical importance of the randomness deviation $\mathcal{D}$ is that it can be brought into a relation with the longest sequence that the given RNG can generate which is still indistinguishable from a true randomness. For a binomial process of bit generation, the bias estimate of a binary sequence of length $N$ is given by:

$$b = -1 + \frac{2}{N}\sum_{i=1}^{N} x_i \qquad (20)$$

with a statistical uncertainty $\sigma_b = 1/\sqrt{N}$, where $N$ is length of the sequence of generated bits $x_i$. We readily have the estimate for autocorrelation, given by Eq. (7), with $\sigma_{a_1} = 1/\sqrt{N}$. Given a sequence of length $N$, one can calculate the statistical uncertainly of $\mathcal{D}$ as follows:

$$\sigma_{\mathcal{D}}^2 = \left(\frac{\partial \mathcal{D}}{\partial a_1}\sigma_{a_1}\right)^2 + \left(\frac{\partial \mathcal{D}}{\partial b}\sigma_b\right)^2 = \frac{2\mathcal{D}}{N\ln(2)}. \qquad (21)$$

Requiring that $\mathcal{D}$ cannot be statistically significantly determined from the sequence, namely that $\mathcal{D} \leq \sigma_{\mathcal{D}}$, one obtains the following criterion on its maximum length:

$$N_{max} = \frac{2}{\ln(2)\mathcal{D}}. \qquad (22)$$

The Eq. (22) is interpreted as follows. A random number generator featuring (i.e. suffering from) a randomness deviation $\mathcal{D}$ can be used to generate a sequence of maximum $N_{max}$ bits which is still undiscernible from a true random sequence. Even though it has been deduced for a Markov process, I hypothesize that it holds generally.

An idealized binomial process which generates bit value of 1 with probability $p = 1/2$ satisfies **D1** and therefore features $\mathcal{D} = 0$. On the other hand, a binomial process having a different but constant probability $p$ will be less random ($\mathcal{D} > 0$). Still, random sequences obtained through such a process are quite useful, for example for stochastic computing [29]. They can be approximated to an arbitrary precision using only generators featuring $p = 1/2$ which therefore can be considered "universal" [24].

**4. Definition of a random bit sequence**

Given **D1**, it is now possible to naturally define a random number sequence.

**Definition D3**. *Random number sequence is a sequence of numbers generated sequentially by a random number generator which conforms to the definition **D1***.

Note that there is no limitation in length of the random sequence: the definition is equally suited for finite and infinite sequences. Unlike in previous definitions, for a random sequence of length $N$, all possible combinations of zeros are equally probable and *can* appear. Thus, the definition endorses a completely unbiased set of output sequences and it does not require that numbers generated thereof pass any set of statistical test as a criterion of compliance to it.

The answer to the historical misconception, mentioned in the Introduction, that one has to exclude "sequences that no one would call random", for example those which contain patterns that are highly compressible [1, 15], such as a sequence of 100 zeros in a row, our answer is that such, or in fact *any* particular output sequence, is random if and only if it has been generated by a correctly operating physical RNG which satisfies the definition **D1**. The definition will take care that each sequence is generated with an exact probability it deserves. For example 100 zeros in a row, albeit possible, will appear with a tiny probability of $2^{-100}$.

Indeed, excluding any sequence(s), only makes the leftover set or the generating process *less random*. An illustration of that is a common practice of some bank card issuers to forbid the use of "risky" 4-digit pins, such as 0000, 1111, ..., 9999, 1234, 4321 etc., not understanding that the public knowledge of this rule makes guessing the PIN by an attacker all that easier due to the reduced set of possibilities that need to be probed.

Furthermore, the definition **D3** supports an important intuitive property of randomness which to our best knowledge has not been mentioned before, nor is supported by any known definition: the *concatenability*. Concatenability requires that concatenation of any number of finite random sequences, either of the identical length or of different lengths, form a random sequence. Mathematical definitions of finite random sequences may not satisfy this very natural requirement because they are based on exclusion of some sequences and those sequences will be completely absent in the concatenated sequence, as already mentioned in section 1.2.

The proof of concatenability of random sequences conforming to **D3** is as follows. Let us have a RNG which conforms to **D1**. Let it generate, one after another, $m$ sequences of lengths $n_1, n_2, ..., n_m$ respectively. Each of the sequences is random by virtue of **D3**. Now, let $N = \sum_{i=1}^{m} n_i$ be the length of the sequence obtained by concatenation of the aforementioned $m$ sequences. This sequence is simply a sequence of $N$ bits being generated by the RNG and therefore is a random sequence as per **D3**. Q.E.D.

## 5. Conclusion

In this work a new definition of randomness has been proposed which, unlike the previous definitions, does not try to define a sequence of random numbers, but instead defines a physical random number generator (RNG) as a device capable of generating a sequence of numbers which is neither self-correlated nor biased. Thus, randomness proof, in practice, consists of proving the two requirements formulated in the definition **D1**.

An exemplary verification of compliance with the **D1** has been demonstrated for three variants of a beamsplitter-based RNG: the variant with a perfectly balanced beamsplitter and flawless single-photon detectors, and two variants with the hardware imperfections most frequently encountered in practice: the beam splitting imbalance and detectors with the dead time.

The fact that imperfect RNGs generate less randomness has motivated the definition of randomness deviation $\mathcal{D}$ (definition **D2**): a quantitative measure whose value spans from 0 for a random process, to 1 for a deterministic (non-random) one. The benefit of it is that once having a robust estimate of $\mathcal{D}$ at hand for a given physical generator, it is not necessary to generate and test long sequences of random numbers generated by it in order to establish their randomness. An approximate expression for $\mathcal{D}$ has been derived under the assumption of a Markov process, which is quite often either a correct or a good approximate model of physical processes used to generate random numbers.

Finally, a definition of a finite random number sequence is presented (**D3**), which is free of problems that plague previous ones, such as arbitrary randomness criteria and concatenability.

I believe that tools developed here will be useful in scientific research, technical innovations and future international standards, regarding random number generators.

## APPENDICES

### A1. Testing randomness tests

In the previous research [24, 25] we constructed a Markov-chain-compliant physical QRNG device with randomness deviation $< 10^{-18}$. Using the criterion of randomness assurance given in Eq. (22), sequences generated by this generator shorter than $N_{max} \approx 3 \cdot 10^{18}$ should be indistinguishable from a truly random sequence by any criterion whatsoever, including any statistical test.

Random data generated by this QRNG enabled us to discover a problem in the FFT test [26] of the NIST STS suite version 2.1.2. [16]. In particular, we have found that bit streams longer than $10^9$ bits result in the FFT test failing too frequently and that for streams longer than $10^{10}$ bits the test fails almost always. Indeed, the existence of that problem has been confirmed in a private communication with NIST in 2019.

This is an example of how a verifiable definition of randomness can be used to "test the tests", as well as to provide much more reliable evidence of randomness than statistical tests alone.

## A2. Randomness of a pseudorandom number generator

Pseudorandom number generators (PRNGs) are frequently used to generate sequences of numbers. An excellent review on PRNG techniques can be found in Ref. [7]. Their main advantage is that they can be implemented as a small computer program, thus not requiring any additional hardware such as a physical RNG connected to the computer. In fact, all computer languages already have a default pseudorandom generator implemented in a form of a function that returns a uniform floating point random variate.

The only downside is that they do not generate random numbers. In order to prove that, let us consider a general pseudorandom generating algorithm in the form [7]:

$$x_{i+1} = F(X_p, S) \qquad (23)$$

where $F$ is a fixed, deterministic generating function describable with a binary sequence $F$, $X_p$ denotes a (sub)set of previously generated numbers, and $S$ is a finite binary constant called the *seed*. Without the loss of generality we take that $x_{i+1}$ is a binary number. By definition, the information content of the process is limited:

$$H(F) + H(S) < N \qquad (24)$$

where $N$ is an integer number. The generating algorithm is carefully chosen such that the expected value $E(x)$ is equal to $1/2$ by design and therefore the requirement Eq. (5) of **D1** is automatically fulfilled. But as I will show, the other requirement may not be satisfied. I will now prove that Eq. (23) does not generate random bits by contradiction (*reductio ad absurdum*).

Proof. Let us assume that Eq. (23) generates random bits. Then, a sequence of the first $i$ generated bits, $X_p$, is a random sequence, By virtue of **D1** the following holds:

$$I(x_{i+1}; X_p) = 0. \qquad (25)$$

Because $x_{i+1}$ is deterministically computable explicitly via Eq. (23), all information about it is contained in $X_p \cup F \cup S$, where $\cup$ signifies concatenation. Taking into account Eq. (25) which states that $X_p$ contains no information on $x_{i+1}$, it follows:

$$I(x_{i+1}; F \cup S) = 1. \qquad (26)$$

Now, let the process of Eq. (23) generate a vector of random bits $x$ whose length is $L > N$. According to Eq. (15) each random bit carries 1 unit of entropy, so its total entropy is

$$H(x) = L. \qquad (27)$$

On the other hand, because $x$ has been deterministically deduced from $F$ and $S$, its entropy must satisfy the following inequality:

$$H(x) \leq H(F) + H(S). \qquad (28)$$

Substituting Eqs. (24) and (27) into Eq. (28) one gets $L \leq N$, which is in contradiction with the initial assumption on lengths, namely $L > N$. This proves that the initial hypothesis, being that Eq. (23) generates random bits, is false. Q.E.D.

In fact, it is straightforward to see that the entropy of a sequence of length $L$, generated by Eq. (23), may not exceed $N$ and therefore its per-bit entropy asymptotically goes to zero with an upper limit of $N/L$.

Thus, any pseudorandom RNG fails Eq. (4) of the definition **D1**. The conclusion of this is that randomness of any deterministic RNG asymptotically tends to zero. Pseudorandom sequences are not even Martin-Löf random either, because they are strongly compressive: by use of Eq. (23) one can reproduce random sequence of any length based on $F$ and $S$ only. This all the more affirms the statement, already mentioned in the introduction, that passing statistical tests does not prove randomness. In fact, we may add now that using statistical tests to evaluate randomness can be highly misleading.

To quote John von Neumann "Anyone who attempts to generate random numbers by deterministic means is, of course, living in a state of sin." [27]. Nevertheless, some pseudorandom generators possess statistical properties good enough to pass virtually all known statistical randomness tests [16], with a small probability to fail in some specific situation [9]. Because of that, PRNGs are frequently used in a variety of applications. Their use is equivalent of having a cheap meal of highly processed junk food against a more expensive but healthy alternative. One has to be warned though that because most PRNGs have been cryptanalyzed, or their short seed may be guessed or attacked, they are generally not advisable for use in cryptography [28].

**Acknowledgments**

**Funding:** This research was funded by Croatian Ministry of Science and Education, grant KK.01.1.1.01.0001.
**Author contributions:** M.S. has conceived and developed the presented theory and wrote the manuscript.

**Competing interests:** Authors declares not to have competing interests.

**Data and materials availability:** Random number sequences and code can be obtained from the author at a reasonable request.